\documentclass{amsart}
\usepackage{graphicx}
\newtheorem{thm}{Theorem}[section]

\theoremstyle{definition}

\theoremstyle{remark}

\numberwithin{equation}{section}
\newcommand{\qudit}[1]{\left\vert #1 \right\rangle}
\newcommand{\rqudit}[1]{\left\langle #1 \right\vert}

\newcommand{\Z}{\mathbb{Z}}

\newcommand{\F}{\mathbb{F}}

\newcommand{\C}{\mathbb{C}}

\usepackage{amssymb}
\font\euler=msbm7  

\usepackage{amssymb, color}

\begin{document}

\title[]{On Deriving a basis for the vector space of bounded qudit error operators  over $\C^d$  }

\author{Colin Wilmott and Peter Wild}
\address{School of Mathematical Sciences\\University College Dublin\\ Dublin 4, Ireland}
\address{Department of Mathematics\\ Royal Holloway\\ University of London\\ Egham\\ Surrey\\ TW20 0EX\\UK}
\thanks{\emph{Electronic address:} cmwilmott@maths.ucd.ie}
\thanks{\emph{Electronic address:} P.Wild@rhul.ac.uk}%

% ----------------------------------------------------------------
\begin{abstract}
We  derive a basis for the vector space of bounded operators
acting on a $d$-dimensional system Hilbert space $\C^d$. In the
context of quantum computation the basis elements are identified
as the generalised Pauli matrices - the error generators. As an
application,  we show how such matrices are used in the
teleportation  a single qudit.
\end{abstract}
\maketitle
% ----------------------------------------------------------------

\section{Introduction}

 The theory of quantum computation continues to advance our understanding of information
 as established in the seminal work of Shannon (1948) through an innovative analysis of the
 nature of noise. This development of a quantum mechanical computing framework has
 redefined quantum computation 
and inspires discoveries whose very nature lie at the frontier of
reality. Constructing a quantum computer is predicated on
realising the inherent  processing advantage of quantum
computation over its classical analogue and  on controlling the
sensitive quantum interference effects that explain the source of
its  computational  power. However, computations are taken in
open quantum systems which  produce unwanted interactions between
sensitive quantum information and noise in the environment. It is
this interaction that results in decoherence - an outcome that
destroys quantum information. Unfortunately,  decoherence is an
inevitable feature of quantum computation, and therefore, it is of
fundamental importance that any coupling between information and
the environment be controlled.  Therefore, to better understand
the fundamentals of noise propagation is to understand the
formalism of  a model that explains it.

In this paper we construct a basis for the space of bounded
operators acting on a $d$-dimensional quantum system $\C^d$. As an
application, we generalise the qubit teleportation scheme.

\section{Preliminaries}
Given an arbitrary finite alphabet $\Sigma$ of cardinality $d$, we
process quantum information by specifying a state description of a
finite dimension quantum space. In particular, the state
description of the Hilbert space $\C^{d}$. While the state of an
$d$-dimensional Hilbert space can be more generally expressed as a
linear combination of  basis states $\qudit{\psi_i}$, we write
each orthonormal basis state of the $d$-dimensional Hilbert space
$\C^{d}$  to correspond with an element of $\Z_{d}$. In this
context the basis $\{\qudit{0}, \qudit{1}, \dots, \qudit{d-1}\}$
is referred to as the \emph{computational basis}. Therefore, a
state $\qudit{\psi}$ of $\C^{d}$ is given by
\begin{eqnarray}
\qudit{\psi} = \sum_{i=0}^{d-1}{}\alpha_i\qudit{i},
\end{eqnarray}
where $\alpha_i \in \C$ and
$\sum_{i=0}^{d-1}{}{\vert\alpha_i\vert}^2 = 1$. A $qudit$
describes a state in the Hilbert space $\C^{d}$. The state space
of an  $n$-qudit state is the tensor product of the basis states
of the single system $\C^{d}$, written ${\mathcal{H}} =
({\C^{d}})^{\otimes n}$, with corresponding orthonormal basis
states given by
\begin{eqnarray}
\qudit{i_1}\otimes\qudit{i_2}\otimes\dots\otimes\qudit{i_n} =
\qudit{i_1}\qudit{i_2}\dots\qudit{i_n} = \qudit{i_1i_2\dots i_n},
\end{eqnarray}
where $i_j \in \Z_{d}$.  The general state of a qudit in the
Hilbert space ${\mathcal{H}}$ is then written
\begin{eqnarray}
\qudit{\psi} = \sum_{(i_1i_2\dots i_n) \ \in \
\textrm{\euler{Z}}_{d}^{n} }{}\alpha_{(i_1i_2\dots
i_n)}\qudit{i_1i_2\dots i_n},
\end{eqnarray} where $\alpha_{(i_1i_2\dots i_n)}  \in  \C$ and $\sum{}\vert\alpha_{(i_1i_2\dots i_n)}\vert^2 =1$.

\section{An Error Model}
The challenge of quantum information processing is to elicit a
reliable form of communication and to maintain such a form in the
presence of quantum noise. Noise is a characteristic of the
{environment} associated with an information state and is a
property of an  {open quantum system} that subjects an information
state to  unwanted  interactions with the elements of the
environment during teleportation. It is inevitable that the
communication of an information state  will cause interactions
with the  environment. However, prolonged contact between the
information state and environment is soon to suffer in
entanglement that degrades the  information state resulting in
decoherence. Any strategy to stabilize quantum computations from
the effects of noise will ultimately be required to  deal with
both  the problems of decoherence and unitary imperfections of
channel communication. To this end, we give the following
description of error within the environment system.

Given a  qudit information state  $\qudit{\psi} =
\sum_{i=0}^{d-1}{}\alpha_i\qudit{i}$   of the Hilbert space
$\C^{d}$, let us consider an adjoined environment space
$\qudit{E}$ endowed with an orthonormal basis of dimension $d^2$.
We suppose that both the state space of the qudit and the
corresponding environment space are initially  independent
systems. The joint state of the systems $\qudit{\psi}$ and
$\qudit{E}$ is then $\qudit{\psi} \otimes \qudit{E}$ and  its
dynamics  may be characterised when we further suppose that the
joint system evolves according to some unitary operation. Given a
unitary operation $U$, we write  interaction of each basis qudit
with the environment  under $U$ as
\begin{eqnarray}
U(\qudit{i} \otimes \qudit{E}) &=&
\sum_{l=0}^{d-1}{}\gamma_{-i+l,-i}(\qudit{i+l}\otimes
\qudit{e_{-i+l,-i}})\nonumber\\&=&
\sum_{l=0}^{d-1}{}\qudit{i+l}\otimes
\gamma_{-i+l,-i}\qudit{e_{-i+l,-i}},\end{eqnarray} for $i \in
\{0,\dots,d-1\}$. By linearity of $U$, the dynamics of the joint
system $\qudit{\psi} \otimes \qudit{E}$ is then
\begin{eqnarray}
U(\qudit{\psi} \otimes
\qudit{E})&=&U\left(\left(\sum_{i=0}^{d-1}{}\alpha_i\qudit{i}\right)\otimes
\qudit{E}\right)\nonumber\\
&=& U\left(\sum_{i=0}^{d-1}{}\alpha_i(\qudit{i}\otimes
\qudit{E})\right)\nonumber\\
&=& \sum_{i=0}^{d-1}{}\alpha_iU(\qudit{i}\otimes
\qudit{E})\nonumber\\
&=& \sum_{i=0}^{d-1}{}\sum_{l=0}^{d-1}{}\alpha_i\qudit{i+l}
\otimes
\gamma_{-i+l,-i}\qudit{e_{-i+l,-i}}.\label{env}\end{eqnarray}
Since $\frac{1}{d}\sum_{z=0}^{d-1}{}\omega^{zk} = 1$ if $z=0$ and
vanishes otherwise then equation (\ref{env}) may be written as
\hskip-2em\begin{eqnarray}
&&\hskip1em \frac{1}{d}\sum_{i=0}^{d-1}{}\sum_{l=0}^{d-1}{}\left(\alpha_i\qudit{i+l} \otimes\left( \sum_{z=0}^{d-1}{}\sum_{k=0}^{d-1}{}\omega^{zk}\gamma_{-i+l+z,-i+z}\qudit{e_{-i+l+z,-i+z}}\right)\right)\nonumber\\
&& \hskip2em = \frac{1}{d}\sum_{i=0}^{d-1}{}\sum_{l=0}^{d-1}{}\sum_{k=0}^{d-1}{}\left(\alpha_i\qudit{i+l}\otimes \left(\sum_{z=0}^{d-1}{}\omega^{zk}\gamma_{-i+l+z,-i+z}\qudit{e_{-i+l+z,-i+z}}\right)\right)\nonumber\\
&& \hskip2em = \frac{1}{d}\sum_{l=0}^{d-1}{}\sum_{k=0}^{d-1}{}\left(\sum_{i=0}^{d-1}{}\left(\alpha_i\qudit{i+l}\otimes \left( \sum_{z=0}^{d-1}{}\omega^{zk}\gamma_{-i+l+z,-i+z}\qudit{e_{-i+l+z,-i+z}}\right)\right)\right)\nonumber\\
&& \hskip2em =
\frac{1}{d}\sum_{l=0}^{d-1}{}\sum_{k=0}^{d-1}{}\left(\sum_{i=0}^{d-1}{}\left(\omega^{ik}\alpha_i\qudit{i+l}\otimes
\left(\sum_{z=0}^{d-1}{}\omega^{-ik}\omega^{zk}\gamma_{-i+l+z,-i+z}\qudit{e_{-i+l+z,-i+z}}\right)\right)\right)\nonumber
\end{eqnarray}\vskip-1em\begin{eqnarray}
&=&
\hskip-.7em\frac{1}{d}\sum_{l=0}^{d-1}{}\sum_{k=0}^{d-1}{}\left(\sum_{i=0}^{d-1}{}\left(\omega^{ik}\alpha_i\qudit{i+l}\otimes
\left(\sum_{z'=0}^{d-1}{}\omega
^{z'k}\gamma_{z'+l,z'}\qudit{e_{z'+l,z'}}\right)\right)\right)\nonumber\\
&=&\hskip-.7em
\frac{1}{d}\sum_{l=0}^{d-1}{}\sum_{k=0}^{d-1}{}\left(\left(\sum_{i=0}^{d-1}{}
\omega^{ik}\alpha_i\qudit{i+l}\right)\otimes
\left(\sum_{z'=0}^{d-1}{}\omega^{z'k}\gamma_{z'+l,z'}\qudit{e_{z'+l,z'}}\right)\right).
\end{eqnarray}

\noindent An  outer product representation describes the set of
operators that act on the joint quantum state under $U$. The
operator $X_1 = \sum_{i=0}^{d-1}{}\qudit{i+1}\rqudit{i}$ maps
$\alpha_i\qudit{i}$ to $\alpha_i\qudit{i+1}$ for $i \in
\{\qudit{0}, \dots,\qudit{d-1}\}$, and thus maps
$\sum_{i=0}^{d-1}{}\alpha_i\qudit{i}$ to
$\sum_{i=0}^{d-1}{}\alpha_i\qudit{i+1}$. Similarly, $Z_1 =
\sum_{i=0}^{d-1}{}\omega^i\qudit{i}\rqudit{i}$ maps
$\alpha_i\qudit{i}$  to $\omega^i\alpha_i\qudit{i}$ and
correspondingly maps $\sum_{i=0}^{d-1}{}\alpha_i\qudit{i}$ to
$\sum_{i=0}^{d-1}{}\omega^i\alpha_i\qudit{i}$. Both $X_1$ and
$Z_1$ are called the \emph{Weyl Pair} (Weyl (1931)). Consequently,
the action of $U$ on $\qudit{\psi}\otimes\qudit{E}$ is described
by the set of operators $X_lZ_k =
\sum_{i=0}^{d-1}{}\omega^{ik}\qudit{i+l}\rqudit{i},$ $ \ (l,k) \in
\Z_{d}\times\Z_{d}$,
\begin{eqnarray}
&&\sum_{l=0}^{d-1}{}\sum_{k=0}^{d-1}{}\left(\left(\sum_{i=0}^{d-1}{}\omega^{ik}\alpha_i\qudit{i+l}\right)\otimes
\frac{1}{d}\left(\sum_{z'=0}^{d-1}{}\omega^{z'k}\gamma_{z'+l,z'}\qudit{e_{z'+l,z'}}\right)\right)\nonumber\\
%\label{construction}
&& = \sum_{l=0}^{d-1}{}\sum_{k=0}^{d-1}{}X_lZ_k\qudit{\psi}
\otimes \gamma_{lk}\qudit{e_{lk}}\label{XZ}.
\end{eqnarray}
Thus,  to correctly specify an  error model that describes  the
action of a unitary operator $U$ on the joint space
$\qudit{\psi}\otimes\qudit{E}$, it is necessary that the
environment $\qudit{E}$, associated  with an information state in
$\C^{d}$, be a Hilbert space of dimension $d^2$. Following the
action of  $U$ on the joint system, a measurement on the
environment is performed  with respect  to the basis
$\qudit{e_{mn}},\ (m,n)  \in \Z_{d}\times\Z_{d}$ to diagnose the
introduced  error in result (\ref{XZ}). Therefore, equation
(\ref{XZ}) provides the conceptual foundation of quantum error
correction. Measurements taken in the environment basis initiate
the correction step $({X_{m}Z_{n}})^{-1} = Z_{(-n\ \textrm{mod}\
d)}X_{(-m\ \textrm{mod}\  d)}$.

 We now  show that the set  $\{X_lZ_K\}, \ (l,k) \in \Z_d \times
 \Z_d,$ $(X_0Z_0 = I)$ constitutes a basis for the space of bounded
 operators on $\C^d$.  As such,  the set $\{X_lZ_K\}, \ (l,k) \in \Z_d \times
 \Z_d,$ forms a $d^2$-dimensional Lie algebra with a $d \times d$
 matrix representation defined by the matrices
with entries {{X}}$_{m,n} = \delta_{m,n-l (\textrm{mod} \ d)}$,
{{Z}}$_{m,n} = \omega^{mk}\delta_{m,n}$.

\begin{thm}
\label{niceproof} Denote by $\omega$ the primitive $d^{th}$ root
of unity. Let us consider $X_{i}\vert k \rangle = \qudit{k + i\
(\textrm{mod}\ {d})}$ and $Z_{j}\qudit{k} = \omega^{kj}\qudit{k}$.
Then ${\mathcal{E}} = \left\{ X_{i}Z_{j} \ \vert \ (i,j) \in
\Z_{d} \times \Z_{d}\right\}$ is a  basis for the space of bounded
 operators  acting on $\C^d$.
\end{thm}
\emph{Proof:} To  show that elements of ${\mathcal{E}}$ are
linearly independent and span $\C^d$, it suffices to show that the
basis $\{\qudit{a}\rqudit{b}\}, \ a,b \in \Z_{d}$, on $\C^d$ may
be expanded as a linear combination of elements in ${\mathcal{E}}$
as both sets of operators are of size $d^2$. Let us consider
${\mathcal{E}}$ in the $\{\qudit{a}\rqudit{b}\}$ basis as
\begin{eqnarray}
E_{i,j} = \sum_{k=0}^{d-1}{}\omega^{jk}\qudit{k+i}\rqudit{k}
\end{eqnarray}
then ${{E}}_{i,j}\qudit{l} = X_iZ_j\qudit{l} =
X_i\omega^{jl}\qudit{l} = \omega^{jl}\qudit{l+i}$. Suppose we may
express $\qudit{a}\rqudit{b}$ as  the linear combination
$\qudit{a}\rqudit{b} = \sum_{(i,j) \in
\Z_{d}\times\Z_{d}}{\mathcal{\xi}}_{i,j}{{E}}_{i,j}$. Then
coefficient ${\xi}_{i,j}$ is    given by
\begin{eqnarray}
{\mathcal{\xi}}_{i,j} &=& \frac{1}{d}\textrm{tr}\left({{E}}_{i,j}^{\dagger}\qudit{a}\rqudit{b}\right)\nonumber\\
&=&\frac{1}{d}\textrm{tr}\left(\sum_{k=0}^{d-1}\omega^{-jk}\qudit{k}\rqudit{k+i}{a}\rangle\rqudit{b}\right)\nonumber\\
&=&\frac{1}{d}\ \omega^{-bj}\  \langle b+i \vert a
\rangle,\nonumber\\
 &=&\frac{1}{d}\ \omega^{-bj}\ \delta_{b+i,a},
\end{eqnarray}
where $\delta_{i,j}$ is the Kronecker delta; \begin{eqnarray}\delta_{i,j} = \left\{%
\begin{array}{ll}
    1, & \hbox{\textrm{if} \ $i = j$} \\
    0, & \hbox{\textrm{if} \ $i \ne j$.}\\
\end{array}%
\right. \end{eqnarray}  We show that with $\xi_{ij}$ defined as
these values then $\qudit{a}\rqudit{b}$ is in the span of
${\mathcal{E}}$. Now,
\begin{eqnarray}
{{E}}^{\dagger}_{i,j}\qudit{a}\rqudit{b} &=&
{{E}}^{\dagger}_{i,j}\sum{}\xi_{k,l}{{E}}_{k,l}\nonumber\\&=&
\xi_{i,j}I + \sum_{k,l\ne
i,j}{}\xi_{k,l}{{E}}^{\dagger}_{i,j}{{E}}_{k,l}
\end{eqnarray}
where $ {{E}}^{\dagger}_{i,j}{{E}}_{k,l}$ has  vanishing trace.
Since
\begin{eqnarray}
\sum_{(i,j) \in
\Z_{d}\times\Z_{d}}{}\frac{1}{d}\omega^{-bj}\delta_{b+i,a}{{E}}_{i,j}
&=& \sum_{(i,j) \in
\Z_{d}\times\Z_{d}}{}\frac{1}{d}\omega^{-bj}\delta_{b+i,a}\left(\sum_{k=0}^{d-1}{}\omega^{jk}\qudit{k+i}\rqudit{k}\right)\nonumber\\
&=& \sum_{k=0}^{d-1}{}\sum_{(i,j) \in \Z_{d}\times\Z_{d}}{}\frac{1}{d}\omega^{(k-b)j}\delta_{b+i,a}\qudit{k+i}\rqudit{k}\nonumber\\
&=& \sum_{k=0}^{d-1}{}\sum_{j=0}^{d-1}{}\frac{1}{d}\omega^{(k-b)j}\qudit{k+a-b}\rqudit{k}\nonumber\\
&=& \qudit{a}\rqudit{b}
\end{eqnarray}
as $\sum_{j}{}\frac{1}{d}\omega^{(k-b)j} = \delta_{k,b}$,  then
$\langle b \vert \sum_{(i,j) \in \Z_{d}\times\Z_{d}} {}
{\xi}_{i,j}{{E}}_{i,j}\vert a \rangle = \delta_{a,b}$. Hence,
$\qudit{a}\rqudit{b} = \sum_{(i,j) \in \Z_{d}\times\Z_{d}}{}
{\xi}_{i,j}{{E}}_{i,j}$ and the result follows.

\section{Quantum Qudit Teleportation}

We now consider the transmission of quantum information with respect to a quantum noisy channel
 where a full continuum of noise is maintained. While classical information can be transmitted
 and protected from the effects of noise by replication, quantum information cannot be copied
 with perfect fidelity (Dieks (1982), Wootters and Zurek
(1982)). Introduced by Bennett \emph{et~al.} (1983), \emph{quantum teleportation} is an
experimental demonstration of the means by which quantum communication is made possible and purports a fundamental distinction between quantum and classical information theory. Such distinction is maintained by the Bell-EPR correlations whereby an essential nonlocality principle, %In physics, the principle of locality is that distant objects cannot have direct influence on one another: an object is influenced directly only by its immediate surroundings.
 described by quantum entanglement, is revealed.
 This result was demonstrated experimentally by Aspect \emph{et
 ~al.} (1982).
 Quantum teleportation takes advantage of the non-local behaviour of quantum mechanics by treating
 quantum entanglement as an information resource. While the Church-Turing Principle maintains
 that it is impossible to transmit quantum information by implementing a classical
computation,  Bennett \emph{et~al.} (1983) introduced quantum
teleportation to overcome this limitation by developing a quantum
algorithm that describes  a complete communication transmission of
quantum information. The first complete transmission of quantum
information  was performed by Nielsen \emph{et~al.} (1998). In the
quantum  teleportation protocol two parties $\mathcal{A}$ and
$\mathcal{B}$  share a pair of particles in a maximally entangled
state. If we suppose that $\mathcal{A}$ is presented with a
quantum  system in an unknown quantum state $\qudit{\psi}$ then
$\mathcal{A}$ can make $\qudit{\psi}$ re-appear at $\mathcal{B}$'s
location (see Fig.~\ref{re-appear}). Central to the protocol is
the use of entanglement to transmit the quantum information of the
unknown state between the parties. We now describe the protocol.
\setlength{\unitlength}{0.08cm} \hspace*{55mm} %\hskip4.5em
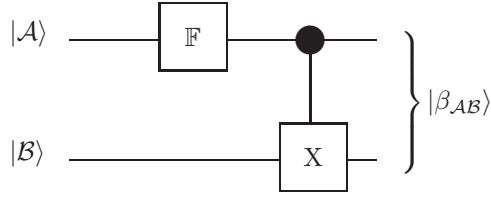
\begin{figure}\label{Bell}
\begin{picture}(70,60)(40,10)
\put(50,45){\line(1,0){15}}\put(76,45){\line(1,0){25}}\put(90,45){\circle*{5}}
\put(90,31){\line(0,1){13}} \put(65,40){\framebox(11,11){$\F$}}
\put(85,20){\framebox(11,11){$\textrm{X}$}}
\put(50,25){\line(1,0){35}}\put(96,25){\line(1,0){5}}
%\put(105,33){$\qudit{\beta_{\textrm{AB}}}$}
\put(105,33.5){$\left\}\begin{matrix} \vspace*{15mm}
\cr\end{matrix}\right.$$\qudit{\beta_{{\mathcal{AB}}}}$}
\put(40,45){$\qudit{\mathcal{A}}$}
\put(40,25){$\qudit{\mathcal{B}}$}
\end{picture}
\caption{Generalised Bell State.}
\end{figure}

Quantum teleportation describes how two parties, $\mathcal{A}$ and
$\mathcal{B}$, process and communicate quantum information in a
manner secure  from the effects of error. Suppose $\mathcal{A}$
wishes to communicate the state $\qudit{\psi}$ then the goal of
teleportation is  to transmit that particular quantum information
state to $\mathcal{B}$. Further suppose that $\mathcal{A}$
prepares the qudit $\qudit{\mathcal{A}}$ where
${\qudit{\mathcal{A}}} \in \{\qudit{0}, \qudit{1}, \dots,
\qudit{d-1}\}$. Similarly, $\mathcal{B}$ prepares the qudit
$\qudit{\mathcal{B}}$ where
 $\qudit{{\mathcal{B}}} \in \{\qudit{0}, \qudit{1}, \dots, \qudit{d-1}\}$.
 In order to achieve teleportation, $\mathcal{A}$ must interact the information
 state $\qudit{\psi}$ with a two qudit entangled state, $\qudit{\beta_{{\mathcal{AB}}}}$.
 The entangled state $\qudit{\beta_{{\mathcal{AB}}}}$ is called  a \emph{generalised Bell state}
  whereby $\mathcal{A}$ and $\mathcal{B}$ each possess one qudit of this two qudit state. To construct
  a  generalised Bell state $\qudit{\beta_{{\mathcal{AB}}}}$, we first apply the  Fourier
  transform $\F\otimes I$ to the qudit $\qudit{A}$. This acts on basis
  states $\qudit{j}\qudit{k}$ as follows $(\F\otimes I)\qudit{j}\qudit{k} =
   \frac{1}{\sqrt{d}}\sum_{i=0}^{d}{}\omega^{ij}\qudit{i}\qudit{k}$
where $\omega$ is a primitive ${d}^{\textrm{th}}$ root of unity in
$\C$ such that $\omega^{d} = 1 $ and $\omega^{t} \ne 1 $ for all
$0<t<d$. Secondly, we follow the Fourier transform by the
controlled-NOT  operation given by $\qudit{k}\qudit{l} \mapsto
\qudit{k}\qudit{l+k\ (\textrm{mod}\ d)}$ for all basis states
$\qudit{k}\qudit{l}$ which maps the  two qudit state accordingly.
Consequently, any pair of qudits
$\qudit{\mathcal{A}}\qudit{\mathcal{B}}$ from the $d^2$
computational basis states of $\C^{d} \otimes \C^{d}$ generate a
generalised Bell state. In particular, applying the Fourier
transform to the first half of the pair of qudit states
$\qudit{\mathcal{A}}\qudit{\mathcal{B}}$, we obtain,
\begin{eqnarray}
&&\hskip-1em \left(\frac{1}{\sqrt{{d}}}\sum^{d-1}_{i=0}{}\sum^{d}_{j=0}{}\sum^{d-1}_{x=0}{}{\omega^{ix}}\qudit{x}\qudit{j}\rqudit{i}\rqudit{j}\right)\qudit{\mathcal{A}}\qudit{\mathcal{B}}\nonumber\\
&& =  \frac{1}{\sqrt{{d}}}\sum^{d-1}_{i=0}{}\sum^{d-1}_{j=0}{}\sum^{d-1}_{x=0}{}{\omega^{ix}}\qudit{x}\qudit{j}\langle{i}\vert{\mathcal{A}}\rangle\langle{j}\vert{\mathcal{B}}\rangle\nonumber\\
&& =
\frac{1}{\sqrt{{d}}}\sum^{d-1}_{x=0}{}{\omega^{{\mathcal{A}}x}}\qudit{x}\qudit{{\mathcal{B}}}\label{Fourier}.\end{eqnarray}
The action of the controlled-NOT operator on  resulting state
(\ref{Fourier}) completes the generalised Bell state construction
\begin{eqnarray}
&&\hskip-2em\left(\sum_{k=0}^{d-1}{}\sum_{l=0}^{d-1}{}\qudit{k}\qudit{l+k}\rqudit{k}\rqudit{l}\right)
\frac{1}{\sqrt{{d}}}\sum^{d-1}_{x=0}{}{\omega^{{\mathcal{A}}x}}\qudit{x}\qudit{{\mathcal{B}}}\nonumber\\
\hskip1em&& = \frac{1}{\sqrt{{d}}}\sum_{l=0}^{d-1}{}\sum_{k=0}^{d-1}{}\sum^{d-1}_{x=0}{}{\omega^{{\mathcal{A}}x}}\qudit{k}\qudit{l+k}\langle{k}\vert x \rangle\langle{l}\vert {\mathcal{B}}\rangle\nonumber\\
&& = \frac{1}{\sqrt{{d}}}\sum^{d-1}_{x=0}{}{\omega^{{\mathcal{A}}x}}\qudit{x}\qudit{{\mathcal{B}}+x} \nonumber\\
&& = \qudit{\beta_{\mathcal{A B}}}\label{betaAB}.
\end{eqnarray}
Since the Bell pair is an entangled state Nielsen and Chuang
(2000), any operator acting on the first qudit held by
$\mathcal{A}$ influences the state of the second qudit held by
$\mathcal{B}$. This condition permits the teleportation of the
quantum information state $\qudit{\psi}$ between parties
$\mathcal{A}$ and $\mathcal{B}$ when $\mathcal{A}$ interacts
$\qudit{\psi}$ with the first half of the generalised Bell pair
(\ref{betaAB}). To negate the effects of the Bell state
transformations  on $\qudit{\psi}$, thereby allowing the
teleportation of $\qudit{\psi}$, $\mathcal{A}$ transforms
$\qudit{\psi}$ by applying the inverse of the generalised
controlled-NOT operator for qudit states which is then followed by
an application of the inverse Fourier transform. Now, the Fourier
transform is unitary so its inverse is its adjoint, and the
inverse of the generalised controlled-NOT operation has its action
defined as $\qudit{k}\qudit{l} \mapsto \qudit{k}\qudit{l-k\
(\textrm{mod}\ d)}$. We write the state of the quantum system held
by $\mathcal{A}$ and $\mathcal{B}$,  as \begin{eqnarray}
\qudit{\psi}\qudit{\beta_{\mathcal{AB}}}&=&\frac{1}{\sqrt{{d}}}\sum_{a=0}^{d-1}{}\alpha_{a}\qudit{a}\left(\sum^{d-1}_{x=0}{}{\omega^{{\mathcal{A}}x}}\qudit{x}\qudit{{\mathcal{B}}+x}\right).
\end{eqnarray}
$\mathcal{A}$ initiates teleportation of the quantum information
state $\qudit{\psi}$ by  applying the inverse  generalised
controlled-NOT operation between $\qudit{\psi}$ and the qudit of
the generalised Bell state held by $\mathcal{A}$, thereby
obtaining,
\begin{eqnarray}
&&\hskip-1em\left(\sum_{k=0}^{d-1}{}\sum_{l=0}^{d-1}{}\sum^{d-1}_{m=0}{}
\qudit{k}\qudit{l-k}\qudit{m}\rqudit{k}\rqudit{l}\rqudit{m}\right)\frac{1}{\sqrt{{d}}}\sum_{a=0}^{d-1}{}
\alpha_{a}\qudit{a}\sum^{d-1}_{x=0}{}{\omega^{{\mathcal{A}}x}}\qudit{x}\qudit{{\mathcal{B}}+x}\nonumber\\
&& = \frac{1}{\sqrt{{d}}}\sum_{k=0}^{d-1}{}\sum_{l=0}^{d-1}{}\sum^{d-1}_{m=0}{}\sum_{a=0}^{d-1}{}\sum^{d-1}_{x=0}{}\alpha_{a}{\omega^{{\mathcal{A}}x}}\qudit{k}\qudit{l-k}\qudit{m}\langle{k}\vert a \rangle\langle{l} \vert x \rangle \langle{m}\vert{{\mathcal{B}}+x}\rangle\nonumber\\
&& =
\frac{1}{\sqrt{{d}}}\sum_{a=0}^{d-1}{}\sum^{d-1}_{x=0}{}\alpha_{a}{\omega^{{\mathcal{A}}x}}\qudit{a}\qudit{x-a}\qudit{{\mathcal{B}}+x}.
\label{outcome}
\end{eqnarray}
\begin{figure}
\setlength{\unitlength}{0.08cm} \hspace*{5mm} \hskip-4.5em
\begin{picture}(15,15)(50,30)
\put(0,0){\line(1,0){107.2}} \put(127.9,0){\line(1,0){5}}
\put(152.5,0){\line(1,0){5}} \put(0,18){\line(1,0){24.9}}
\put(90,35){\line(1,0){50}} \put(90,37){\line(1,0){50}}
\put(141,5){\line(0,1){30}} \put(143,5){\line(0,1){30}}
\put(35,18){\line(1,0){45}} \put(90,17){\line(1,0){27}}
\put(90,19){\line(1,0){27}} \put(119,5){\line(0,1){15}}
\put(117,5){\line(0,1){15}} \put(60,36){\line(1,0){20}}
\put(0,36){\line(1,0){50}} \put(25,13){\framebox(10,10){$X^{-1}$}}
\put(30,36){\circle*{3}} \put(142,36){\circle*{5}}
\put(118,18){\circle*{5}} \put(30,23){\line(0,1){13}}
\put(50,31){\framebox(10,10){$\F$}}
\put(80,13){\framebox(10,10){$M_2$}}
\put(80,31){\framebox(10,10){$M_1$}}
\put(107.5,-5){\framebox(20,10){$X_{-{\mathcal{B}}-M_2}$}}
\put(132.5,-5){\framebox(20,10){${Z_{-{\mathcal{A}}-M_1}}$}}
\put(-18,7.5){$\qudit{\beta_{\mathcal{AB}}}$
$\left\{\begin{matrix} \vspace*{13mm} \cr\end{matrix}\right.$}
\put(-14,35.5){${\qudit{\psi}}$} \put(165,-1){${\qudit{\psi}}$}
\end{picture}
\vskip8em\caption{Quantum channel for teleporting a
qudit.}\label{re-appear}
\end{figure}
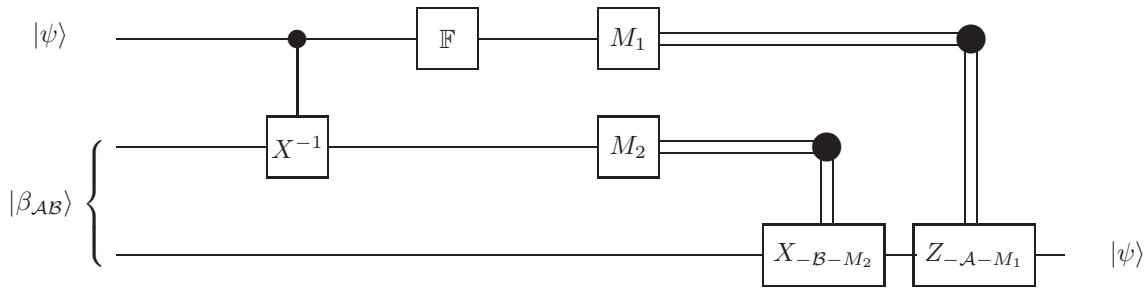Following this result, $\mathcal{A}$ applies the discrete Fourier transformation on the
first qudit of the state (\ref{outcome}). The outcome of this operation is to place the
state (\ref{outcome}) into the state given by
\begin{eqnarray}
&&\left(\frac{1}{\sqrt{{d}}}\sum_{i=0}^{d-1}{}\sum_{y=0}^{d-1}{}\sum_{j=0}^{d-1}{}\sum_{n=0}^{d-1}{}\omega^{iy}
\qudit{y}\qudit{j}\qudit{n}\rqudit{i}\rqudit{j}\rqudit{n}\right)\frac{1}{\sqrt{{d}}}\sum_{a=0}^{d-1}{}\sum^{d-1}
_{x=0}{}\alpha_{a}{\omega^{{\mathcal{A}}x}}\qudit{a}\qudit{x-a}\qudit{{\mathcal{B}}+x}\nonumber\end{eqnarray}\begin{eqnarray}
&& = \frac{1}{{{d}}}\sum_{i=0}^{d-1}{}\sum_{y=0}^{d-1}{}\sum_{j=0}^{d-1}{}\sum_{n=0}^{d-1}{}\sum_{a=0}^{d-1}{}\sum^{d-1}_{x=0}{}\alpha_{a}\omega^{iy}{\omega^{{\mathcal{A}}x}}\qudit{y}\qudit{j}\qudit{n}\langle{i}\vert a \rangle\langle{j}\vert x-a\rangle\langle{n}\vert {\mathcal{B}}+x\rangle\nonumber\\
&& =
\frac{1}{{{d}}}\sum_{y=0}^{d-1}{}\sum_{a=0}^{d-1}{}\sum^{d-1}_{x=0}{}\alpha_{a}\omega^{ay}{\omega^{{\mathcal{A}}x}}\qudit{y}\qudit{x-a}\qudit{{\mathcal{B}}+x}\nonumber\\
&& = \frac{1}{{{d}}}\sum_{y=0}^{d-1}{}\sum_{a=0}^{d-1}{}\sum^{d-1}_{x=0}{}\sum_{z=0}^{d-1}{}\alpha_{a}\omega^{ay}{\omega^{{\mathcal{A}}x}}\qudit{y}\qudit{z}\langle{z}\vert {x-a}\rangle\qudit{{\mathcal{B}}+x}\nonumber\\
&&
 = \frac{1}{{{d}}}\sum_{y=0}^{d-1}{}\sum_{a=0}^{d-1}{}\sum_{z=0}^{d-1}{}\alpha_{a}\omega^{ay}{\omega^{{\mathcal{A}}(z+a)}}\qudit{y}\qudit{z}\qudit{{\mathcal{B}}+z+a}\nonumber
\\
&& =
\frac{1}{{{d}}}\sum_{y=0}^{d-1}{}\sum_{z=0}^{d-1}{}\omega^{{\mathcal{A}}z}\qudit{y}\qudit{z}\left(\sum_{a=0}^{d-1}{}\alpha_{a}\omega^{a(y+{\mathcal{A}})}\qudit{{\mathcal{B}}+z+a}\right).\label{tele}
\end{eqnarray}
The qudit of the generalised Bell state held by  $\mathcal{B}$ is transformed into the state $\sum_{a=0}^{d-1}{}\alpha_{a}\omega^{a(y+{\mathcal{A}})}\qudit{{\mathcal{B}}+z+a}$. Thus $\mathcal{A}$ has teleported a quantum information state $\qudit{\psi'}$ to $\mathcal{B}$, however, it has been subjected to \emph{error} over the channel and therefore $\mathcal{B}$ receives  $\sum_{a=0}^{d-1}{}\alpha_{a}\omega^{a(y+{\mathcal{A}})}\qudit{{\mathcal{B}}+z+a}$ instead of $\sum_{a=0}^{d-1}{}\alpha_{a}\qudit{a}$. A measurement projection onto the computational basis state of $\C^{d} \otimes \C^{d}$ is performed by $\mathcal{A}$ on the first and second qudit of the state of the quantum system (\ref{tele}) which yields two classical numbers. Simultaneously, the third qudit of the  state of the system (\ref{tele})  teleported to $\mathcal{B}$ collapses to a post-measurement state that is dependent upon the measurement outcome obtained by $\mathcal{A}$. Let $M_1,M_2$ be two classical numbers corresponding to the  resulting states $\qudit{M_1}\qudit{M_2}$. %constitute  the projective measurements outcomes recorded by $\mathcal{A}$.
Then the state of the qudit held by $\mathcal{B}$ is given by
$\sum_{a=0}^{d-1}{}\alpha_{a}\omega^{({\mathcal{A}}+M_1)a}\qudit{{\mathcal{B}}+M_2+a}$.
The set $M_1,M_2$ is transferred by classical means to
$\mathcal{B}$, where upon delivery $\mathcal{B}$ learns which of
the generalised Pauli operators are required to correct the effect
of the error. In particular, $\mathcal{B}$ applies the operators
$$X_{-{\mathcal{B}}-M_2} =
\sum_{x=0}^{d-1}{}\qudit{x-{\mathcal{B}}-M_2}\rqudit{x}$$ and
$$Z_{-{\mathcal{A}}-M_1} =
\sum_{z=0}^{d-1}{}\omega^{(-{\mathcal{A}}-M_1)z}\qudit{z}\rqudit{z}$$
in order to return the post-measurement state
$\sum_{a=0}^{d-1}{}\alpha_{a}\omega^{({\mathcal{A}}+M_1)a}\qudit{{\mathcal{B}}+M_2+a}$
to the initial quantum information state $\qudit{\psi}$.  Hence,
applying  $X_{-{\mathcal{B}}-M_2}$ to the post-measurement state,
$\mathcal{B}$  obtains
\begin{eqnarray}
&&X_{-{\mathcal{B}}-M_2}\left(\sum_{a=0}^{d-1}{}\alpha_{a}\omega^{({\mathcal{A}}+M_1)a}\qudit
{{\mathcal{B}}+M_2+a}\right)\nonumber\\
&&
=\sum_{x=0}^{d-1}{}\qudit{x-{\mathcal{B}}-M_2}\rqudit{x}\left(\sum_{a=0}^{d-1}{}\alpha_{a}
\omega^{({\mathcal{A}}+M_1)a}\qudit{{\mathcal{B}}+M_2+a}\right)\nonumber\end{eqnarray}\begin{eqnarray}
&& =\sum_{x=0}^{d-1}{}\sum_{a=0}^{d-1}{}\alpha_{a}\omega^{({\mathcal{A}}+M_1)a}\qudit{x-{\mathcal{B}}-M_2}\langle x \vert {{\mathcal{B}}+M_2+a} \rangle\nonumber\\
&&
=\sum_{a=0}^{d-1}{}\alpha_{a}\omega^{({\mathcal{A}}+M_1)a}\qudit{a}\label{X}.\end{eqnarray}
The operator $Z_{-{\mathcal{A}}-M_1}$ is then applied by
$\mathcal{B}$ on result (\ref{X}) returning  the post-measurement
state to the quantum information state initially held by
$\mathcal{A}$,
\begin{eqnarray}
&&Z_{-{\mathcal{A}}-M_1}\left(\sum_{a=0}^{d-1}{}\alpha_{a}\omega^{({\mathcal{A}}+M_1)a}\qudit{a}\right)\nonumber\\
&&=\sum_{z=0}^{d-1}{}\omega^{(-{\mathcal{A}}-M_1)z}\qudit{z}\rqudit{z}\left(\sum_{a=0}^{d-1}{}\alpha_{a}\omega^{(M_1+{\mathcal{A}})a}\qudit{a}\right)\nonumber\\
&&=\sum_{z=0}^{d-1}{}\sum_{a=0}^{d-1}{}\alpha_{a}\omega^{({\mathcal{A}}+M_1)a}\omega^{(-{\mathcal{A}}-M_1)z}\qudit{z}\langle{z}\vert{a}\rangle\nonumber\\
&&=\sum_{a=0}^{d-1}{}\alpha_{a}\qudit{a}.
\end{eqnarray}
This  ends the teleportation protocol - the state of
$\mathcal{B}$'s  system is left in the same state as the one
initially presented to $\mathcal{A}$. The quantum information can
only be obtained if it vanishes from $\mathcal{A}$ thereby
upholding the no-cloning theorem (Dieks (1982), Wootters and Zurek
(1982)). Thus $\mathcal{B}$ obtains the quantum information which
$\mathcal{A}$ wished to transmit. This is what it means for the
quantum information to have been transmitted (Timpson (2006)).

% ----------------------------------------------------------------

\end{document}